\documentclass[aps, prd, twocolumn, showpacs, floatfix, letterpaper,
 nofootinbib, superscriptaddress,longbibliography]{revtex4-2}

\usepackage{blindtext}
\usepackage[newitem]{paralist}
\usepackage{graphicx}
\usepackage{epsfig}
\usepackage{bm}
\usepackage{amsfonts}
\usepackage{pgf}
\usepackage{color}
\usepackage[T1]{fontenc}
\usepackage[latin1]{inputenc}
\usepackage{graphicx}
\usepackage[english]{babel}
\usepackage{amsmath}
\usepackage{amssymb}
\usepackage{amsfonts}
\usepackage{makecell}
\usepackage{hyperref}
\usepackage[draft,deletedmarkup=xout]{changes}
\usepackage{mathtools}
\usepackage{xcolor}
\usepackage{multirow}
\usepackage{mhchem}
\usepackage{wrapfig}
\usepackage{lipsum}
\usepackage{xcolor}
\usepackage{multirow}
\usepackage{mhchem}
\usepackage{float}
\definecolor{green}{rgb}{0.19,0.64,0.54}
\definecolor{blue}{rgb}{0,0,1}
\definecolor{reddish}{rgb}{0.65, 0.2, 0.2}
\definecolor{darkgreen}{rgb}{0.2,0.7,0.3}
\definecolor{darkblue}{rgb}{0.3,0.40,0.48}
\definecolor{gray}{rgb}{.8,.8,.8}

\hypersetup{,
	pdftitle={CMB from wiggly strings},
  	pdfauthor={authors},
    colorlinks=true,       
    linkcolor=darkblue,          
    citecolor=darkgreen,        
    filecolor=reddish,      
    urlcolor=blue,          
    linktocpage=true
}

\begin{document}

\title{Cosmic Microwave Background anisotropies generated by cosmic strings with small-scale structure}

\author{R. P. Silva}
\email[]{rpdasilva17@gmail.com}
\affiliation{Centro de Astrof\'{\i}sica da Universidade do Porto, Rua das
Estrelas, 4150-762 Porto, Portugal}
\affiliation{Instituto de Astrof\'{\i}sica e Ci\^encias do Espa\c co,
CAUP, Rua das Estrelas, 4150-762 Porto, Portugal}
\affiliation{Departamento de F\'{\i}sica e Astronomia, Faculdade de Ci\^encias, Universidade do Porto, Rua do Campo Alegre 687, PT4169-007 Porto, Portugal}

\author{L. Sousa}
\email[]{lara.sousa@astro.up.pt}
\affiliation{Centro de Astrof\'{\i}sica da Universidade do Porto, Rua das
Estrelas, 4150-762 Porto, Portugal}
\affiliation{Instituto de Astrof\'{\i}sica e Ci\^encias do Espa\c co,
CAUP, Rua das Estrelas, 4150-762 Porto, Portugal}
\affiliation{Departamento de F\'{\i}sica e Astronomia, Faculdade de Ci\^encias, Universidade do Porto, Rua do Campo Alegre 687, PT4169-007 Porto, Portugal}

\author{I. Yu. Rybak}
\email[]{ivan.rybak@astro.up.pt}
\affiliation{Centro de Astrof\'{\i}sica da Universidade do Porto, Rua das
Estrelas, 4150-762 Porto, Portugal}
\affiliation{Instituto de Astrof\'{\i}sica e Ci\^encias do Espa\c co,
CAUP, Rua das Estrelas, 4150-762 Porto, Portugal}

\begin{abstract}

We study the impact of kinks on the cosmic microwave background (CMB) anisotropies generated by cosmic string networks. To do so, we extend the Unconnected Segment Model to describe the stress-energy tensor of a network of cosmic strings with kinks and implement this extension in CMBACT to compute the CMB anisotropies generated by these wiggly string networks. Our results show that the inclusion of kinks leads, in general, to an enhancement of the temperature and polarization angular power spectra, when compared to those generated by cosmic string networks without small-scale structure with the same energy density, on scales corresponding to the distance between kinks. This enhancement, that is more prominent in the case of the temperature anisotropies, is essentially caused by a significant increase of the vector-mode anisotropies, since kinks, due to their shape, generate vortical motions of matter --- a phenomenon that is not taken into account when resorting to an effective description of wiggly cosmic strings.

\end{abstract}

\date{\today}
\maketitle

\section{Introduction}

In several high-energy scenarios, phase transitions may give rise to $1+1$-dimensional topological defects known as cosmic strings in the early universe~\cite{Kibble,Vilenkin:2000jqa,HindmarshKibble}. These strings are not expected to decay quickly and they may then form a network that survives through cosmic history, potentially leading to a variety of observational imprints. In particular, lensing~\cite{Sazhin1,DaisukeYamauchi_2014,Xiao:2022hkl}, gravitational wave~\cite{Auclair:2019wcv} and Cosmic Microwave Background (CMB)~\cite{Lazanu:2014eya,CharnockAvgoustidisCopelandMoss} probes may be sensitive to cosmic string signatures and may enable us to investigate whether cosmic strings have been created in the early universe. Cosmic strings then provide us with the means to probe the symmetry-breaking mechanisms predicted in different high-energy models. Since a variety of scenarios may lead to cosmic strings of different types~\cite{JeannerotRocherSakellariadou, SarangiTye, Dror:2019syi}, understanding their properties is essential to precisely pinpoint the energy scale of the particular phase transition that originates them.

Gravitational waves probe the existence of standard cosmic strings with higher sensitivity than other methods~\cite{BLANCOPILLADO2018392, Ringeval_2017, Auclair:2019wcv}, but these constraints may need to be revised for non-vanilla cosmic strings~\cite{PhysRevD.94.063529, Auclair:2022ylu, RybakSousa2}. At the same time, CMB data provides a complementary independent constraint on the string mass per unit length ($G \mu_0 \lesssim 10^{-7}$~\cite{Planck:2013mgr,Lazanu:2014eya} for standard strings) and may also help to reveal potential signatures of topological defects with additional degrees of freedom~\cite{Rybak:2017yfu, HindmarshLizarragaUrrestillaDaverioKunz}. However, to derive accurate constraints on cosmic string scenarios using CMB data, we need to perform accurate predictions of the perturbations they generate throughout cosmological history. This is often achieved by resorting to the Unconnected Segment Model (USM)~\cite{Albrecht:1997mz, Pogosian:1999np}, which provides a description of the stress-energy tensor of a cosmic string network by treating it as a collection of uncorrelated straight segments. Cosmic strings, however, collide and intercommute often throughout their evolution and this may lead to a build up of small-scale structure on the strings. In this study, we extend the USM, realized in the publicly available CMBACT code~\cite{Pogosian:1999np}, to predict the CMB anisotropies generated by a network of cosmic strings with small-scale structure (also referred to as wiggly strings). Specifically, we substitute the straight segments by zig-zag segments representing cosmic strings with kinks. We then include the effect of small-scale structure by directly changing the shape of string segments in the code. This approach, by relying on a straightforward implementation of kinks into CMBACT, allows us to study directly their impact on the CMB anisotropies, which we then compare to the predictions obtained with an effective description of wiggles~\cite{Pogosian:1999np,Martins:2014kda}, wherein details about the small-scale structure are integrated out. 

This paper is organized as follows. In Sec.~\ref{sec:SEtensor}, we compute the stress-energy tensor of a cosmic string segment with an arbitrary (odd) number of kinks. In Sec.~\ref{sec:model}, we review the USM (Sec.~\ref{subsec:usm}), used to describe the stress-energy tensor of the network, and the Velocity-dependent One-scale (VOS) model (Sec.~\ref{subsec:vos}), which provides a description of the large-scale dynamics of the network, and describe how these are adapted to provide a description of wiggly string networks. We then study the CMB anisotropies generated by a wiggly cosmic string network in Sec.~\ref{sec:anisotropies} and study the impact of kink sharpness --- in Sec.~\ref{subsec:sharp} --- and of the number of kinks per segment --- in Sec.~\ref{subsec:kink} --- on this signature. We compare our results with the results obtained using an effective description of small-scale structure in Sec.~\ref{subsec:comp}. We then discuss our results and conclude in Sec.~\ref{Conc}.

\section{The Stress-energy tensor of a cosmic string with kinks}\label{sec:SEtensor}

\begin{figure}
    \centering
    \includegraphics[scale = 0.55]{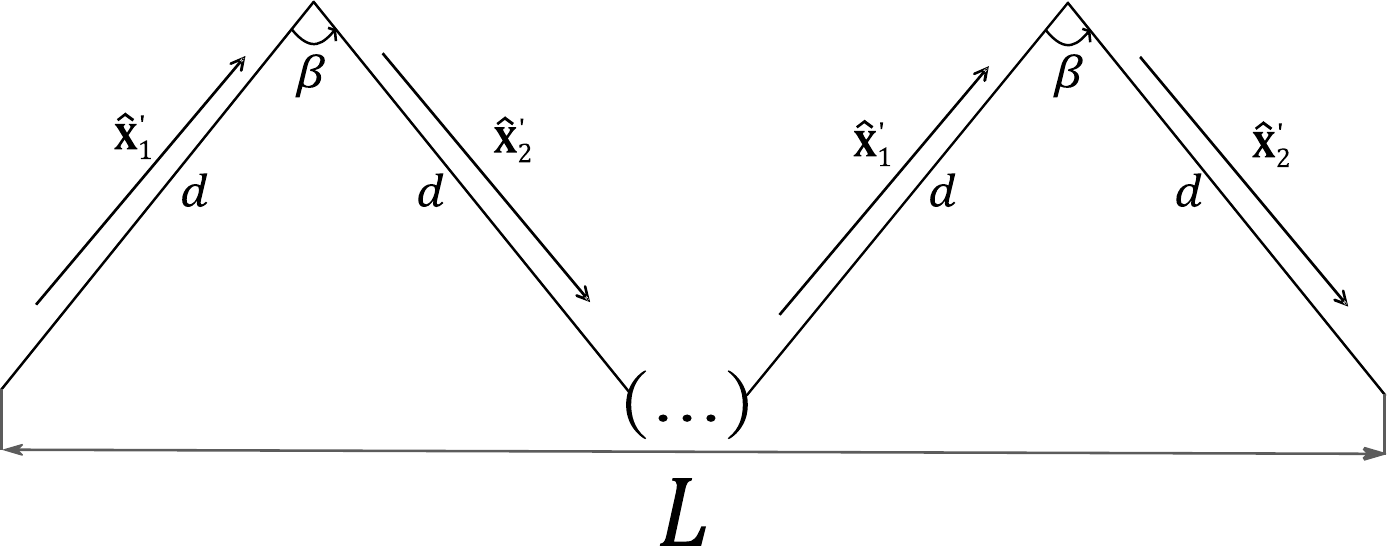}
    \caption{Illustration of a string with $N$ kinks. $L$ is the projective length and $d$ corresponds to the subsegment length.}
    \label{fig:N-kink}
\end{figure}

In their evolution, cosmic strings often collide and interact. When this happens, strings exchange partners and intercommute, which leads to the production of discontinuities known as \textit{kinks}. These frequent intercommutations are expected to lead to a build up of small-scale structure on cosmic strings, making strings wiggly. In this section, we compute the stress-energy tensor of a cosmic string with kinks as a first step to study the impact of small-scale structure on the CMB signatures of cosmic strings.

In most situations of interest in cosmology, cosmic strings may be treated as infinitely thin objects that sweep a $1+1$-dimensional surface in spacetime known as a worldsheet. This worldsheet,
\begin{equation}
X^{\mu} = X^{\mu}(\xi, \sigma) \; ,
\end{equation}
may be parametrized by a timelike coordinate, $\xi$, and a spacetime coordinate, $\sigma$, and its evolution may be described by the Nambu-Goto action
\begin{equation}
\label{NG1}
S = - \mu_{0} \int d\xi d\sigma \sqrt{-\gamma},
\end{equation}
where $\mu_ 0$ is the cosmic string mass per unit length, $\gamma$ is the determinant of the worldsheet metric, $\gamma_{ab} = g_{\mu \nu} X^\mu_{,a} X^\nu_{,b}$  and  $ g_{\mu \nu}$ is the spacetime metric. 

The Nambu-Goto action is invariant under worldsheet reparametrizations, and, in a Friedmann-Lemaitre-Robertson-Walker (FLRW) universe --- whose line element is given by
\begin{equation}
\label{metric}
ds^2 = a(\tau)^2 \left( d \tau^2 - d \textbf{x}^2 \right)\,,
\end{equation}
where $a(\tau)$ is the cosmological scale factor, $d\tau=dt/a$ is the conformal time and $t$ is the physical time --- it is common to chose temporal-transverse gauge conditions

\begin{equation}
\xi = \tau  \; \; \text{and} \; \;   \dot{\mathbf{X}} \cdot \mathbf{X}^{\prime} = 0 \; ,
\end{equation}
where $X^{\mu} = \left(\tau, \mathbf{X} \right)$ and a dot and a prime represent a derivative with respect to $\tau$ or $\sigma$, respectively. In this gauge, $\dot{\mathbf{X}}$ corresponds to the physical velocity of the string and it is perpendicular to the string tangent $\mathbf{X}^{\prime}$. Also, in this case, the stress-energy tensor (obtained by varying Eq. (\ref{NG1}) with respect to $g_{\mu \nu}$) is of the form:
\begin{equation}
T^{\mu \nu} = \frac{\mu_0}{\sqrt{-g}} \int  d \tau d \sigma \left( \epsilon \dot{X}^{\mu} \dot{X}^{\nu} - \epsilon^{-1} X^{\prime \, \mu} X^{\prime \, \nu} \right) \delta^{(4)} \,,
\end{equation}
where $\delta^{(4)} = \delta^{(4)}\left(x^{\sigma} - X^{\sigma}(\tau, \sigma) \right)$ is a Dirac-delta function and $\epsilon^{2} = \mathbf{X}^{\prime \, 2}/\left(1 - \dot{\mathbf{X}}^2\right)$.

To model wiggly strings, we consider cosmic string segments with $N$ kinks that are piecewise straight and have a projective length (or physical extension) of $L$. In other words, we will assume that the string is composed of $N + 1$ straight sub-segments\footnote{Although technically, at the kinks, strings are cannot be considered infinitely thin --- since kinks correspond to highly curved regions --- the Nambu-Goto action is expected to provide an adequate description of the dynamics of these straight subsegments.} with the same length, $d$, that meet at the kinks at an angle $\beta$, as illustrated in Fig.~\ref{fig:N-kink}. The string tangent, then, is discontinuous at the kinks, where the different subsegments meet. For a segment with projective length $L$, the subsegment length $d$ depends both on the number of kinks $N$ and on $\beta$:
\begin{equation}
\label{d eq}
d = \frac{L}{\left(N + 1 \right) \sin\left(\beta/2 \right)} \,,
\end{equation}
and gives us the distance between kinks. Moreover, we shall assume that the string segments are planar and contained in the $xy$-plane and that they behave as a rigid body --- i.e., the velocity is the same at every point of the string --- with a velocity, $\dot{\mathbf{{X}}}$, orthogonal to the $xy$-plane. This velocity, then, should be interpreted as the translational velocity of the segments. Moreover, for simplicity, we assume that kinks do not travel along the strings and we also neglect the impact of oscillations of the strings induced by the high curvature at the kink. Although kinks are quite generally expected to propagate along the string and to induce oscillations at small-scales, these impacts may be included dynamically in the future (as discussed in~\ref{Conc}).

This string configuration may be simply regarded as a series of straight string subsegments disposed sequentially. Thus, the trajectory of each subsegment is given by:

\begin{equation}
    \mathbf{X}_i =  \mathbf{y}_{k} + v \tau \mathbf{\hat{\dot{X}}} + \sigma \hat{  \mathbf{X}}^{\prime}_i \,  \, , 
\end{equation}
where $\mathbf{y}_{k}$ is the position of the $k$-th kink, $k=i,i-1$, depending on whether $i$ is odd or even, respectively, and $\hat{\dot{\mathbf{{X}}}}=\hat{\mathbf{e}}_{z}$ is a unit vector along the direction of the velocity of the string. Moreover, $\hat{\mathbf{X}}^{\prime}_{i} =  \sin(\beta/2)\hat{\mathbf{e}}_{x} \pm \cos(\beta/2)\hat{\mathbf{e}}_{y}$ --- with the plus and minus signs corresponding to odd and even $i$ respectively --- is a unitary vector with the direction of the tangent of string subsegment $i$. Here, $\hat{\mathbf{e}}_{x}$, $\hat{\mathbf{e}}_{y}$ and $\hat{\mathbf{e}}_{z}$ are three orthonormal vectors defined as
\begin{equation}
\label{eq:basis1}
\hat{\mathbf{e}}_{x} =  \begin{pmatrix}
\sin(\theta)\sin(\phi)  \\
-\sin(\theta)\cos(\phi)  \\
\cos(\theta)
\end{pmatrix},
\end{equation}
\begin{equation}
\hat{\mathbf{e}}_{y} =  \begin{pmatrix}
-\cos(\phi)\sin(\psi) - \cos(\psi)\sin(\phi)\cos(\theta)  \\
-\sin(\phi)\sin(\psi) + \cos(\psi)\cos(\phi)\cos(\theta)  \\
\sin(\theta)\cos(\psi)
\end{pmatrix},
\end{equation}
\begin{equation}
\label{eq:basis3}
\hat{\mathbf{e}}_{z} =  \begin{pmatrix}
\cos(\phi)\cos(\psi) - \sin(\psi)\sin(\phi)\cos(\theta)  \\
\sin(\phi)\cos(\psi) + \sin(\psi)\cos(\phi)\cos(\theta)  \\
\sin(\theta)\sin(\psi)
\end{pmatrix},
\end{equation}
where $0 \leq \theta < \pi$ and  $0 \leq \phi, \, \psi < 2 \pi$.

The Fourier transform of the stress-energy tensor of a string segment with $N$ kinks is then given by
\begin{widetext}
\begin{equation}
\label{N kink}
\begin{split}
\Theta^{\mu \nu} = \mu_{0}  \sum_{m=0}^{\left( N-1 \right)/2}\Big[&\int_{-d}^{0} d\sigma e^{i \mathbf{k} \cdot \mathbf{X}_{2m+1}}\left( \epsilon\dot{X}^{\mu}_{2m+1}\dot{X}^{\nu}_{2m+1}- \epsilon^{-1} X_{2m+1}^{\prime \; \mu}  X_{2m+1}^{\prime \; \nu} \right)  
\\
+   &\int_{0}^{d} d\sigma e^{i \mathbf{k} \cdot \mathbf{X}_{2m+2}}\left( \epsilon\dot{X}^{\mu}_{2m+2}\dot{X}^{\nu} _{2m+2}- \epsilon^{-1} X_{2m+2}^{\prime \; \mu}  X_{2m+2}^{\prime \; \nu} \right)    \Big] \,,
\end{split}
\end{equation}
\end{widetext}
where, for simplicity, we have assumed that $N$ is odd (so that we are always considering an even number of subsegments).

The real part of the temporal component of the stress-energy tensor (\ref{N kink}) may then be written as
\begin{widetext}

\begin{equation}
\label{T00}
\begin{split}
\Theta^{00} = &\mu_{0} \gamma \frac{\sin\left(  \frac{L}{2} \mathbf{k} \cdot \mathbf{\hat{e}}_{x}   \right)}{\sin\left(  \frac{L}{N+1} \mathbf{k} \cdot \mathbf{\hat{e}}_{x}   \right)}\Bigg\{ \cos \left( A_{mp} \right) \Bigg[\frac{\sin\left(dk{\hat{X}^{\prime}}_{13}  \right)}{k{\hat{X}^{\prime}}_{13}}   + \frac{\sin\left(dk{\hat{X}^{\prime}}_{23}  \right)}{k{\hat{X}^{\prime}}_{23}}     \Bigg] +  \\
&+ \sin \left( A_{mp} \right)\Bigg[\frac{{\hat{X}^{\prime}}_{23} - {\hat{X}^{\prime}}_{13} }{k{\hat{X}^{\prime}}_{13}{\hat{X}^{\prime}}_{23}}   + \frac{\cos\left(dk{\hat{X}^{\prime}}_{23} \right)}{k{\hat{X}^{\prime}}_{23}}  - \frac{\cos\left(dk{\hat{X}^{\prime}}_{13}  \right)}{k{\hat{X}^{\prime}}_{13}}    \Bigg]\Bigg\} \; , 
\end{split} \; 
\end{equation}
where
\begin{equation}
A_{mp}=
\begin{cases}
\mathbf{k} \cdot \mathbf{y_{mp}}  + kv \tau {\hat{\dot{X}}_{3}} \, , \quad \text{if} \quad \frac{N-1}{2} \,\,\mbox{is even},\\
\mathbf{k} \cdot \mathbf{y_{mp}}   + kv \tau {\hat{\dot{X}}_{3}}  + d \cos(\beta/2) \mathbf{k} \cdot \mathbf{\hat{e}}_{y}\, , \quad \text{if} \quad \frac{N-1}{2} \,\,\mbox{is odd} \,,
\end{cases}
\end{equation}
\end{widetext}
$\mathbf{y_{mp}}$ is the position of the $(N+1)/2$-th kink (located midway along the string) and we have assumed, without loss of generality, that $\mathbf{k} = k \textbf{k}_{3}$, with $\textbf{k}_{3} = \left( 0, \; 0, \; 1 \right)$, and where we defined ${\hat{\dot{X}}_{3}}=\mathbf{\hat{\dot{X}}}\cdot \mathbf{k}_{3}$ and ${\hat{X}^{\prime}}_{i3}=\mathbf{\hat{X}^{\prime}}_{i}\cdot \mathbf{k}_{3}$. Notice that, by setting $\beta=\pi$, we recover the stress-energy tensor of a straight cosmic string derived in~\cite{Albrecht:1997mz,Pogosian:1999np}. Note also that, since this expression does not involve an explicit summation over the subsegments, $N$ can be increased arbitrarily at no additional computational cost. For more details regarding the derivation of Eq.~(\ref{T00}), see~\cite{teserodrigo}.

The spatial components of the stress-energy tensor can be written as
\begin{equation}
\label{Tij}
\Theta^{ij} = \sum_{m = 1}^{2}\left[ v^{2}{{\hat{\dot{X}}}^i_{m}}{{\hat{\dot{X}}}^j_{m}} - \left(1 - v^{2} \right){\hat{{X}}^{\prime \, i}}_{m} {\hat{{X}}^{\prime \, j}}_{m} \right] \Theta^{00}_{(m)} \, ,
\end{equation}
where we have separated the contributions of the odd and even subsegments, labeled respectively by $m=1$ and $m=2$, so that $\Theta^{00}_{(m)}$ includes only the terms in Eq.~(\ref{T00}) involving $\hat{{X}}^{\prime }_{m3}$.

The scalar, vector, and tensor components of the stress-energy tensor (\ref{T00}) are given, respectively, by:
\begin{equation}
\begin{split}
&\Theta_{S} = \left(2\Theta^{33} - \Theta^{11}  - \Theta^{22} \right)/2  \, ,
\\
&\Theta_{V} = \Theta_{V}^{13} \, ,
\\
&\Theta_{T} = \Theta_{T}^{12}    \; .
\end{split} 
\end{equation}
By imposing local energy-momentum conservation \cite{Albrecht:1997mz}, the trace $\Theta = \Theta_{ii}$ and the velocity field $\Theta^{D} = \Theta_{03}$ are fixed.

\section{Modeling a network of wiggly strings}\label{sec:model}

To study the CMB anisotropies generated by wiggly cosmic strings, we will resort to CMBACT~\cite{Pogosian:1999np, Pogosian:2006hg}, a publicly available numerical tool based on CMBFAST~\cite{Seljak:1996is} that was designed to compute the CMB angular power spectra generated by active sources of perturbations. This tool solves the linearized Einstein-Boltzmann equations to compute these anisotropies and so requires a description of the stress-energy tensor of the wiggly cosmic string network for the entire simulation volume to act as a source. For standard cosmic strings, this is often achieved by resorting to the Unconnected Segment Model (USM)~\cite{Albrecht:1997mz,Pogosian:1999np}, that --- by simplifying the network as a collection of uncorrelated and randomly distributed segments --- allows us to construct the stress-energy tensor for the network using that of a single segment. This model is then used alongside the Velocity-dependent One-Scale (VOS) model~\cite{Martins:1996jp,Martins:2000cs}, which provides a thermodynamical description of the large-scale dynamics of the network, to ensure an accurate description of the energy density of the network throughout cosmic history. In this section, we review these models and describe how they may be adapted to describe networks of strings with kinks.

\subsection{The Unconnected Segment Model}\label{subsec:usm}

The USM approximates a network of cosmic strings by a collection of uncorrelated straight string segments with the same (comoving) length $L$ that are simultaneously created at some early time. These segments are assumed to be randomly distributed and orientated and to move in random directions with the same velocity $v$. To mimic the energy-loss experienced by the cosmic string network throughout its evolution, a fraction of the segments decay at each time-step. Although realistic cosmic string networks are expected to be significantly more complex, this framework nevertheless allows us accurately characterize the CMB anisotropies they generate, by averaging the CMB anisotropies generated by a large number of realizations of these very simplified networks. As a matter of fact, the angular power spectrum obtained using this framework is in agreement with that computed using Nambu-Goto simulations (see e.g.~\cite{Planck:2013mgr,Lazanu:2014eya}) and may be calibrated to mimic the results of Abelian-Higgs simulations~\cite{Planck:2013mgr}. Here, we will then maintain the essential features of this approach and simply substitute the straight segments by segments with kinks like those described in the previous section.

Using this simplified framework, the stress-energy tensor of the cosmic string network may be written as the sum of the contributions of each cosmic string~\cite{Pogosian:1999np}:
\begin{equation} 
\label{History Stress-Energy Tensor}
\tilde{\Theta}_{\mu\nu} ( \vec{k} ,\tau ) =
\sum_{m=1}^{N_{s}} \Theta^m_{\mu\nu}  ( \vec{k} ,\tau ) 
T^{\rm off} \left(\tau,\tau^{m}_f\right) \; \; ,
\end{equation} 
where $N_{s}$ is the number of strings contained within the simulation volume and function $T^{\rm off}$ was introduced to model the decay of the $m$-th segment at a conformal time $\tau^{m}_f$. This function may be written as~\cite{Albrecht:1997mz}:
\begin{equation}
\label{Toff}
T^{\text{off}} (\tau,\tau_f)=   
\begin{cases}
    1\,,  &\mbox{if}\,\,  \tau < \lambda_f \tau_f \\
    \frac{1}{2} + \frac{1}{4} \left(x_{\text{off}}^3 - 3 x_{\text{off}} \right)\,,  &\mbox{if} \,\,\lambda_f \tau_f \le \tau < \tau_f \,,\\
    0\,, & \mbox{if}\,\,\tau_f\le\tau
\end{cases}
\end{equation}
where 
\begin{equation}
\label{xoff}
x_{\text{off}} = 2 \frac{ \ln(\lambda_{f} \tau_f / \tau) }{\ln (\lambda_{f})} - 1 \,.
\end{equation}

Note that, in general, one expects to have a large number of string segments in the simulation volume and directly dealing with each individual segment would be computationally costly. To avoid this, all the string segments that decay at the same time may be consolidated into a single segment~\cite{Pogosian:1999np, Albrecht:1997mz}. The number segments $N_{d}$ that decay at a given time $\tau$ is~\cite{Pogosian:1999np, Albrecht:1997mz}:
\begin{equation}
\label{Number of decaying}
N_{d} \left( \tau \right) = V \left[n(\tau_{i -1}) - n(\tau_{i}) \right] \; \; , 
\end{equation}
where $V$ is the simulation volume and $n(\tau)$ is the string number density at time $\tau$. The number density of segments is given by
\begin{equation}
\label{USM1}
n(\tau) = \frac{C(\tau)}{L^{3}} \, \, ,
\end{equation}
where $C(\tau)$ is determined by requiring the total number of strings at any time is given by $V_{0}/L(\tau) ^{3}$ \cite{Pogosian:1999np} and $V_{0}$ is the comoving volume (we will discuss this in more detail in the next subsection). Since randomly oriented individual segments have random phases in Fourier space, the amplitude of the Fourier transformed stress-energy tensor describing the sum of their contributions is essentially that of a single segment weighted by a factor of $\sqrt{N_d}$~\cite{Pogosian:1999np, Albrecht:1997mz, Rybak:2021scp}. Thus, one can write~\cite{Pogosian:1999np}:
\begin{equation}
\label{Stress-Tensor Consolidated}
\tilde{\Theta}_{\mu \nu} (\textbf{k}, \tau) =  \sum_i \sqrt{N_{d}(\tau_i)}\Theta_{\mu\nu}^i(\textbf{k},\tau) T^{\rm off}(\tau,\tau_{i})\,,
\end{equation}
where the index $i$ runs over the consolidated segments.

\subsection{The velocity-dependent One-scale model}\label{subsec:vos}

Note that the USM tells us nothing about the number of strings that exist, their average length and velocity and thus, to have an adequate description of the stress-energy tensor of a realistic cosmic string network, we also need a model to describe these quantities throughout cosmic history. For standard cosmic strings, this is achieved by resorting to the VOS model. This model provides a description of the evolution of the (comoving) characteristic length $L_c$ of the network --- defined in terms of the energy density of the network as
\begin{equation}
    \rho=\frac{\mu_0}{a^2L_c^2}
    \label{eq:char-len}
\end{equation}
--- and of the root-mean-squared (RMS) velocity of the network $v$. For featureless cosmic strings, without significant small-scale structure, the characteristic length $L_c$ is also an appropriate measure of the (averaged) length of segments $L$ and of the (averaged) interstring distance. In other words, a single lengthscale is sufficient to describe the large-scale dynamics of the network and $L$ and $L_c$ may be used interchangeably. In this case, the cosmological evolution of the cosmic string network may be described by~\cite{Martins:1996jp,Martins:2000cs}:
\begin{eqnarray}
\label{Evolution of L}
\frac{dv}{d\tau} & = & \left(1 - v^{2} \right) \left[ \frac{k(v)}{L} - 2 v \frac{\dot{a}}{a} \right]\,\label{eq:vosv}\\
\frac{d L}{d\tau} & = & \frac{\dot{a}}{a} L v^{2} + \frac{c}{2} v\label{eq:vosL}\,,
\end{eqnarray}
where we express the equations in terms of $L$ instead of $L_c$ (as in the original version of the model) since $L\sim L_c$. The last term in~(\ref{eq:vosL}) describes the energy loss caused by the production and subsequent decay of cosmic string loops in collisions and self-intersection of cosmic strings and $c$ is a phenomenological parameter describing the efficiency of this energy loss mechanism. Moreover, $k(v)$ is a momentum parameter that, to some extent, describes the average curvature of cosmic strings. Nambu-Goto numerical simulations are well described by 
\begin{equation}
k(v) \equiv \frac{2\sqrt{2}}{\pi}(1-v^{2})(1+2\sqrt{2}v^{3})\frac{1 - 8v^{6}}{1+8v^{6}}
\end{equation}
and $c=0.23$~\cite{Martins:2000cs}. For strings without small-scale structure, then, Eqs.~(\ref{eq:vosv}) and~(\ref{eq:vosL}) are used to set the number density of string segments throughout cosmic history using Eq.~(\ref{USM1}), thus ensuring that the decay of segments accurately mimics the energy loss caused by loop production and decay. Moreover, it is generally assumed that the bare length --- corresponding to the physical extension of the segments, without the small-scale structure --- and velocity are the same for all segments and given, respectively, by $L$ and $v$.

When the cosmic string segments are wiggly, however, the characteristic length $L_c$ and the length of segments $L$ can therefore be significantly different, which means that the VOS model cannot be directly applied. To circumvent this, here we will assume that the VOS model accurately describes the \textit{bare} energy density of a network of strings, i. e. the energy density excluding the contribution of small-scale structure. The bare energy density of the network $\rho_0$ may be related to the bare length of segment as follows
\begin{equation}
    \rho_0=\frac{\mu_0}{a^2L^2}\,.
\end{equation}
So, in other words, we will always consider strings with the same bare length even if the number of kinks and $\beta$ changes, and we will assume that this length is well described by the VOS model. This is a first approximation to describe the dynamics of networks of wiggly cosmic strings, other more complex models for describing networks of strings with small-scale structure were proposed in~\cite{Vilenkin:1990mz,Carter:1994zs,Austin:1993,Martins:2014kda}. Although here we use the simplest possible model as a first approximation, the adapted USM model for wiggly strings developed here can be used with these more complex models in a very straightforward way in the future (see Sec.~\ref{Conc}).

Note that the total length of each segment is, for wiggly strings, larger than their bare length:
\begin{equation}
    L_{\rm total}=\frac{L}{\sin{\left(\beta/2\right)}}\equiv \mathcal{S} L\,,
\end{equation}
where we have introduced, for succinctness, a \textit{sharpness parameter} $\mathcal{S}$. As the kinks become ``pointier'' --- or, as $\beta$ decreases --- the sharpness of the kinks increases and so does the total length of the segments. This means that the energy of each segment then also increases by a factor of $\mathcal{S}$ when strings become wiggly. Since, by fixing the bare length of segments, we are also fixing the number density of segments, this necessarily means that the energy density of the network also increases. As a matter of fact, we have that:

\begin{equation}
    \rho=\frac{\mu_0 \mathcal{S}}{a^2 L^2}\equiv \frac{\mu_{\rm eff}}{a^2L^2}
    \label{ew:en-density}
\end{equation}
where we have introduced the effective mass per unit length of the cosmic strings
\begin{equation}
    \mu_{\rm eff}=\mu_0\mathcal{S}\,,
    \label{eq:mueff}
\end{equation}
which is larger than $\mu_0$ for $\beta<\pi$ (i.e. if the strings are wiggly). Note that the total length of segments is, in this case, independent of the number of kinks per segment and so the effective mass per unit length and the characteristic length of the network depend only on string sharpness. From Eq.~(\ref{eq:char-len}), we may also see that these wiggly cosmic strings have then a characteristic length that may, in fact, be significantly smaller than the bare length:
\begin{equation}
    L_c^2=L^2/\mathcal{S}\,.
\end{equation}
This clearly shows that the characteristic length cannot be assumed to be an appropriate measure of the projective length of the wiggly segments, as it may be significantly different from $L$.

\section{Cosmic microwave background anisotropies generated by wiggly strings}\label{sec:anisotropies}

The CMB anisotropies are often characterized in terms of the angular power spectrum, $C_\ell$, of its temperature fluctuations:
\begin{equation}
C_{\ell} \equiv \frac{1}{2 \ell + 1} \sum^{\ell}_{m = -\ell} \langle a^{*}_{\ell m} a_{\ell m}  \rangle \, ,
\end{equation}
where $\langle \cdots \rangle$ represents the ensemble average. Here, $a_{lm}$ are the coefficients of the decomposition of the temperature anisotropies in terms of spherical harmonics $Y_{\ell m} (\hat{\mathbf{r}})$,
\begin{equation}
\frac{\Delta T}{T} (\hat{\mathbf{r}}) = \sum_{\ell = 0}^{\infty}\sum_{m = - \ell}^{\ell} a_{\ell m} Y_{\ell m} (\hat{\mathbf{r}}) \, ,
\end{equation}
and $\hat{\mathbf{r}}$ is the direction along the line of sight.

Besides computing the CMB anisotropies generated by wiggly cosmic string networks, we will also compute the linear Cold Dark Matter (CDM) power spectrum of perturbations generated by them, defined as
\begin{equation}
P(k) = \left| \delta^2(\mathbf{k})\right| \, ,
\end{equation}
where $\delta(\mathbf{k})$ is the Fourier transform of the density contrast
\begin{equation}
\delta(\mathbf{x}) = \frac{\rho_{m}(\mathbf{x}) - \langle \rho_{m} \rangle}{\langle \rho_{m} \rangle} \, ,
\end{equation}
where $\rho_{m}$ is the CDM density at the position $\mathbf{x}$ and $\langle \rho_{m} \rangle$ is its average value.

In this section, we will compute the CMB angular power spectrum and the linear CDM power spectra generated by wiggly cosmic strings for different $N$ and $\beta$, with the objective of studying the potential impact of small-scale structure. To do this, we implement the changes described in the previous sections to the publicly available CMBACT code to allow for the description of wiggly cosmic string segments, thus substituting the straight string segments by `zig-zag' segments. Our results are obtained by averaging over 500 realizations of wiggly cosmic string networks --- in which the (consolidated) segments are randomly distributed and oriented (by a random choice of $\mathbf{y_{mp}}$ and of the angles $\theta$, $\phi$ and $\psi$ in Eqs.~(\ref{eq:basis1})-(\ref{eq:basis3})) --- and we take the cosmological parameters measured by the Planck mission~\cite{Planck:2018vyg}. We also fix the effective mass per unit length of the cosmic string networks to $G\mu_{\rm eff}=10^{-7}$, so that, even when we change the sharpness of kinks, the networks have the same energy density.

\subsection{Impact of kink sharpness}\label{subsec:sharp}

Let us start by considering a fixed number of kinks and varying their sharpness to study the impact of wiggliness. As discussed, increasing kink sharpness leads to an increase of the energy density of the cosmic string network and this necessarily leads to an increase of the amplitude of CMB anisotropies. As a matter of fact, one expects roughly that~\cite{Pogosian:2007gi}:

\begin{equation}
    C_\ell \propto \left(G\mu_{\rm eff}\right)^2 \propto \mathcal{S}^2\,.
    \label{eq:mueff-sca}
\end{equation}
Here, since we fix $G\mu_{\rm eff}=10^{-7}$, we shall investigate whether making segments wiggly causes any deviations from this behaviour. Note that, when we replace straight segments by `zig-zag' segments, we are effectively introducing an additional scale to the problem: the subsegment length or interkink distance $d$. This naturally leads to changes to the spectrum of perturbations generated by the strings.

\begin{figure}[h!]
    \centering
    \includegraphics[width=3.4in]{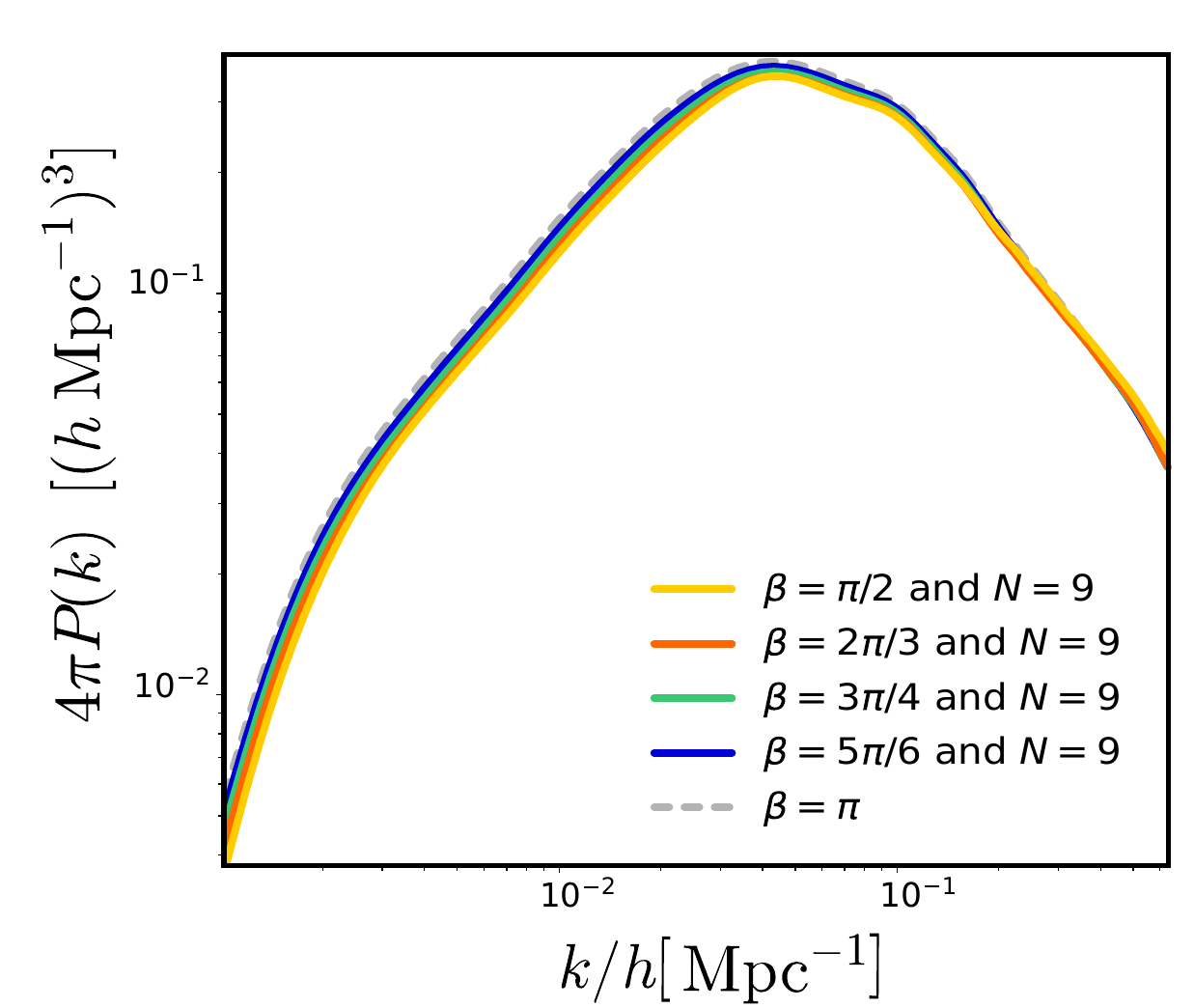}
    \caption{CDM Linear power spectrum generated by wiggly cosmic strings for different values of $\beta\ge\pi/2$ and $N=9$. The dashed line corresponds to the power spectrum generated by straight cosmic strings. Here we set $G\mu_{\rm eff}=10^{-7}$ and the results are obtained by averaging over 500 realizations.}
    \label{fig:PSlarge}
\end{figure}
\begin{figure}[h!]
    \centering
    \includegraphics[width=3.4in]{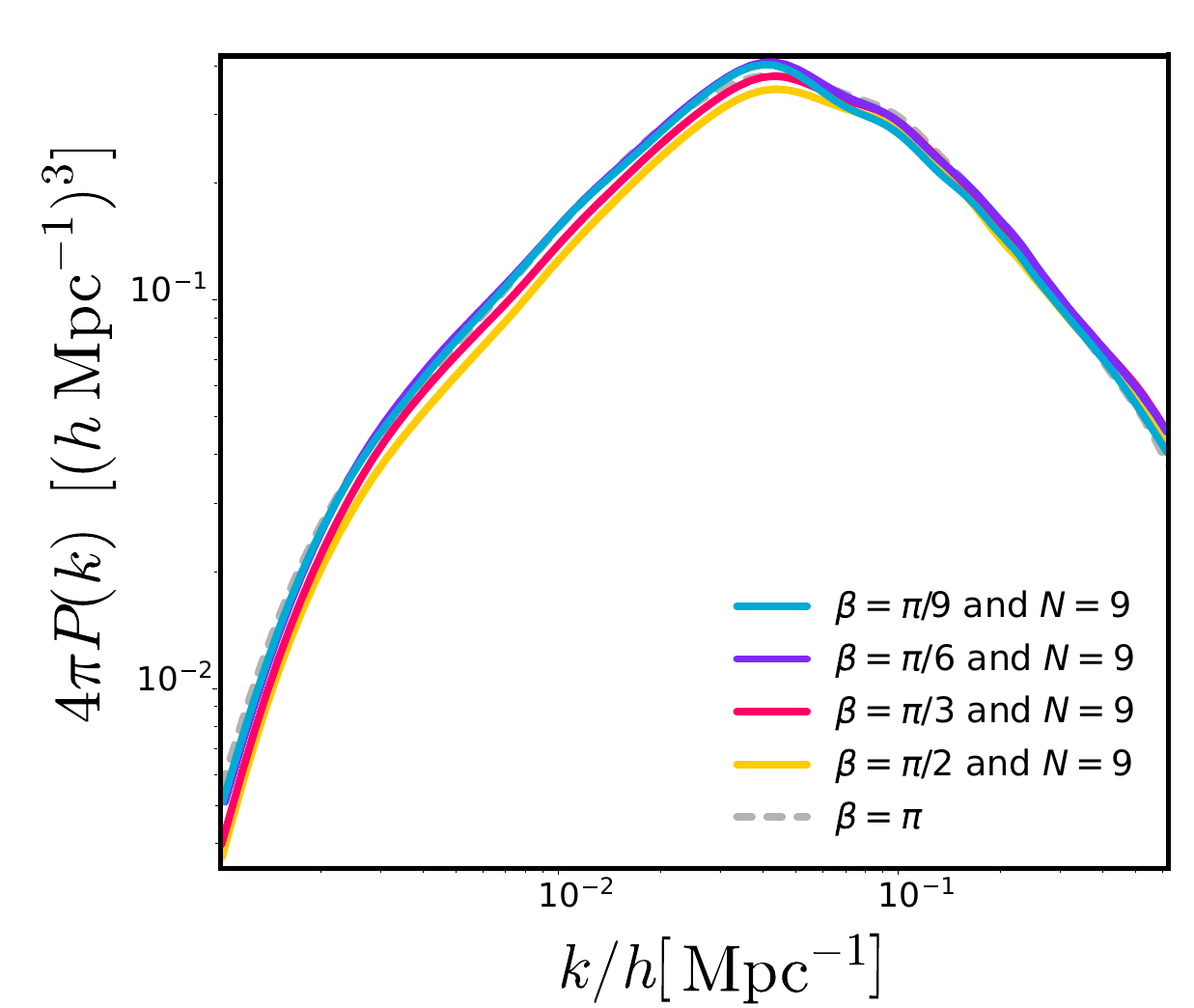}
    \caption{CDM Linear power spectrum generated by wiggly cosmic strings for different values of $\beta\le\pi/2$ and $N=9$. The dashed line corresponds to the power spectrum generated by straight cosmic strings. Here we set $G\mu_{\rm eff}=10^{-7}$ and the results are obtained by averaging over 500 realizations.}
    \label{fig:PSsmall}
\end{figure}

\begin{widetext}

\begin{figure}[h!] 
            \centering
            \includegraphics[width=7in]{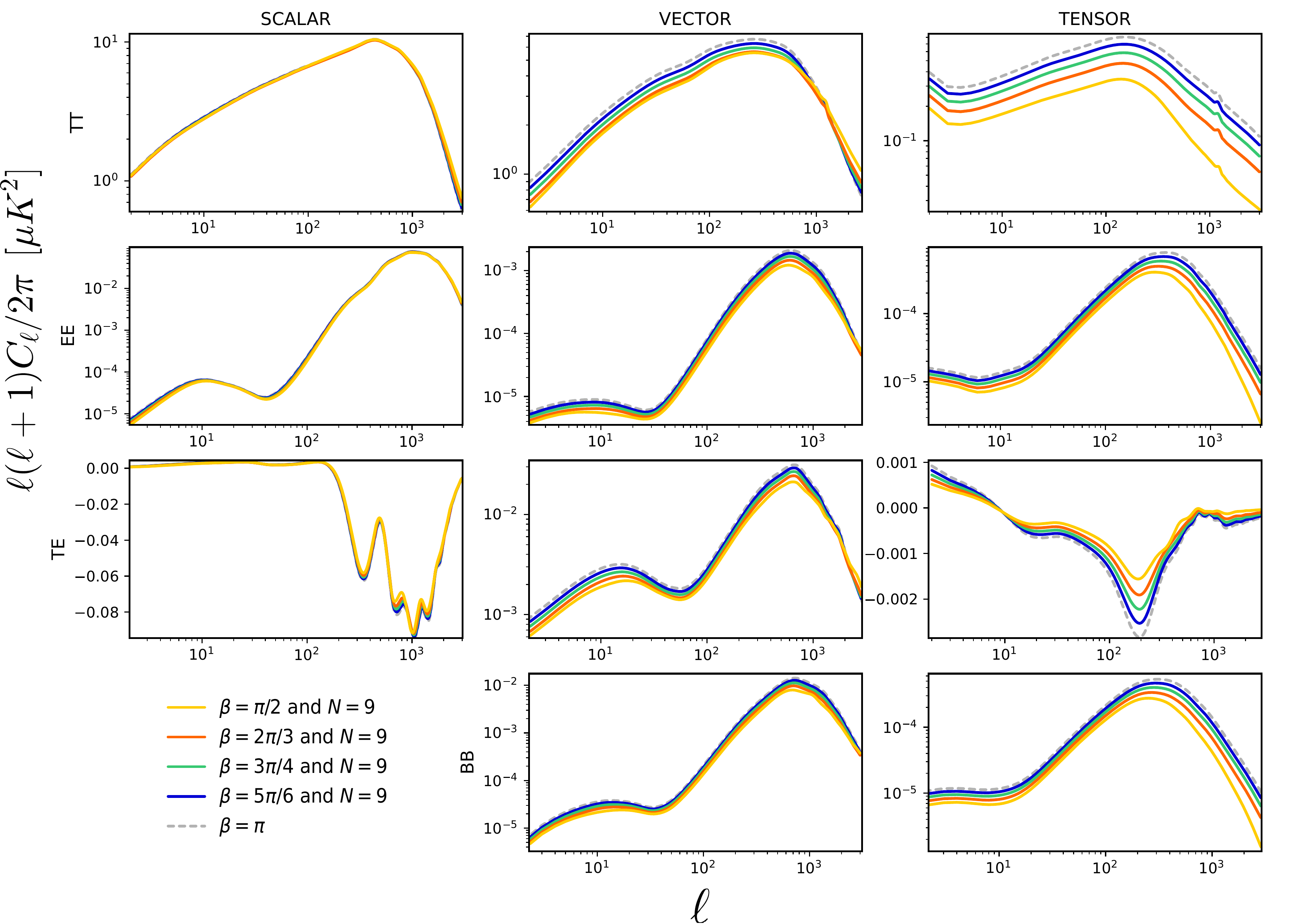}
            \caption{CMB anisotropies generated by wiggly cosmic strings for different values of $\beta\ge\pi/2$ and $N=9$. From top to bottom, we plot the TT, EE, TE and BB power spectra, as a function of the multipole moment $\ell$. The left, middle and right columns represent the scalar, vector and tensor components, respectively. The dashed lines correspond to the angular power spectra generated by straight cosmic strings. Here we set $G\mu_{\rm eff}=10^{-7}$ and the results are obtained by averaging over 500 realizations.}
            \label{fig:CMBlarge}
\end{figure}    

\begin{figure}[h!] 
            \centering
            \includegraphics[width=7in]{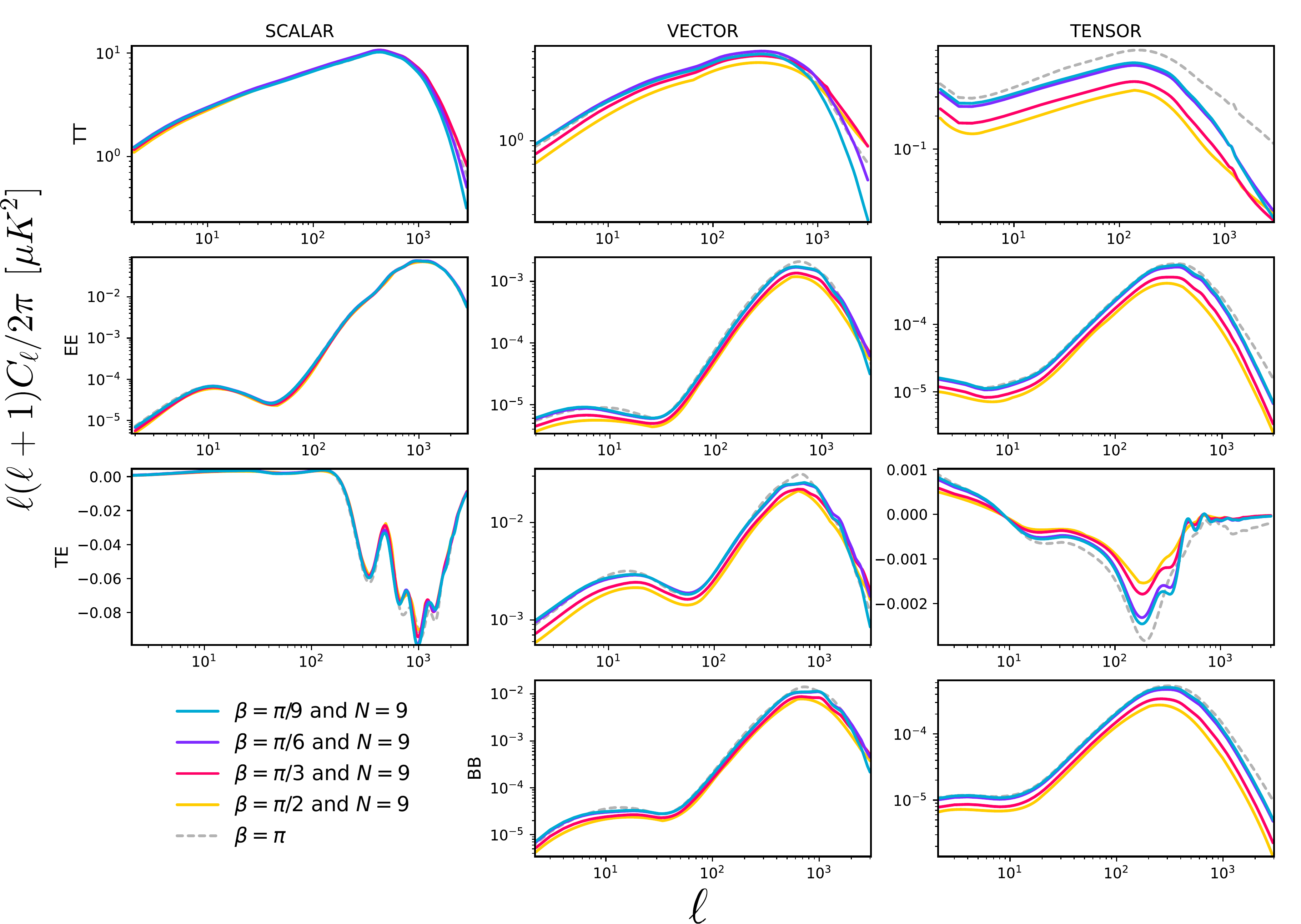}
            \caption{CMB anisotropies generated by wiggly cosmic strings for different values of $\beta\le\pi/2$ and $N=9$. From top to bottom, we plot the TT, EE, TE and BB power spectra, as a function of the multipole moment $\ell$. The left, middle and right columns represent the scalar, vector and tensor components, respectively. The dashed lines correspond to the angular power spectra generated by straight cosmic strings. Here we set $G\mu_{\rm eff}=10^{-7}$ and the results are obtained by averaging over 500 realizations.}
            \label{fig:CMBsmall} 
\end{figure}    

\end{widetext}

This behaviour may clearly be seen in Figs.~\ref{fig:PSlarge} and~\ref{fig:PSsmall}, where we plot the CDM linear power spectrum generated by networks of strings with different values of kink sharpness. As Fig.~\ref{fig:PSlarge} --- where this spectrum is plotted for $\beta\ge \pi/2$ --- illustrates, as we make strings wiggly, there is a shift of power from large to small-scales. Note that the spectrum's peak is roughly at a scale corresponding to the length of the segments~\cite{Wu:1998mr} and, in this limit, $d \lesssim 0.1 L$, which explains this enhancement for large $k$. However, if we increase the sharpness of kinks further, since $d\propto \mathcal{S}$, the enhancement of the CDM linear power spectrum happens on progressively larger scales (or smaller $k$). This results in a higher peak height, as may be seen in Fig.~\ref{fig:PSsmall}.

These changes to the spectrum of perturbations generated by strings naturally translate into changes to the angular CMB power spectrum they generate, as Figs.~\ref{fig:CMBlarge} and~\ref{fig:CMBsmall} show. In Fig.~\ref{fig:CMBlarge}, where the scalar, vector and tensor components of the temperature anisotropies, E-mode and B-mode polarization power spectra and the TE correlation are plotted for $\beta\ge\pi/2$, one may see that the scalar components are not significantly affected by changing kink sharpness. As was the case for the linear CDM power spectrum, we also see a shift of power from large to small-scales, which is more prominent for the temperature anisotropies. When we consider even sharper kinks, with $\beta\le\pi/2$, --- as is the case for the angular power spectra plotted in Fig.~\ref{fig:CMBsmall} --- the picture is not significantly changed. The scalar components are still not significantly affected, except for the fact that there is a shift of power towards the scale corresponding to the subsegment length. This means that, as the kinks become very sharp, there is a slight enhancement of the scalar TT anisotropies on large scales.

Vector and tensor components, however, are significantly affected by the inclusion of kinks. In both cases, there is an overall decrease of the amplitude of the anisotropies for $\beta\ge \pi/2$, which is in general more prominent for the tensor modes. We also see an enhancement of the vector components on small-scales that, for the TT anisotropies, is much more prominent. However, as we further decrease $\beta$, there is a reversal of this trend: both vector and tensor anisotropies start to increase (but, for very small $\beta$, this increase starts to decelerate significantly). In the case of the vector component of the temperature anisotropies, this increase may be quite significant and the amplitude of this component may even exceed that of straight strings over most of the spectrum. Pointier kinks, then, due to their geometry, result in vortical motions of matter, leading to an enhancement of vector modes. For large $\ell$, however, as for scalar anisotropies, there is a decrease of the amplitude of the spectrum as $\mathcal{S}$ grows and this effect may be quite significant for very sharp kinks, resulting in a significant narrowing of the peak of the spectrum. The same suppression of power may be seen in the tensor component, but much less prominent.

As to the E-mode and B-mode polarization, there is also a decrease of the amplitude of the vector and tensor components of the power spectra for small enough $\beta$ and, for $\beta \gtrsim \pi/2$, we also see an increase of their amplitude (except on small enough scales). Overall, the shape of the corresponding angular power spectra is quite similar to that generated by strings without small-scale structure --- albeit with a smaller amplitude --- but the peaks have, quite generally, a smaller relative height.

\subsection{Impact of the number of kinks}\label{subsec:kink}

\begin{figure}[h!]
    \centering
    \includegraphics[width=3.4in]{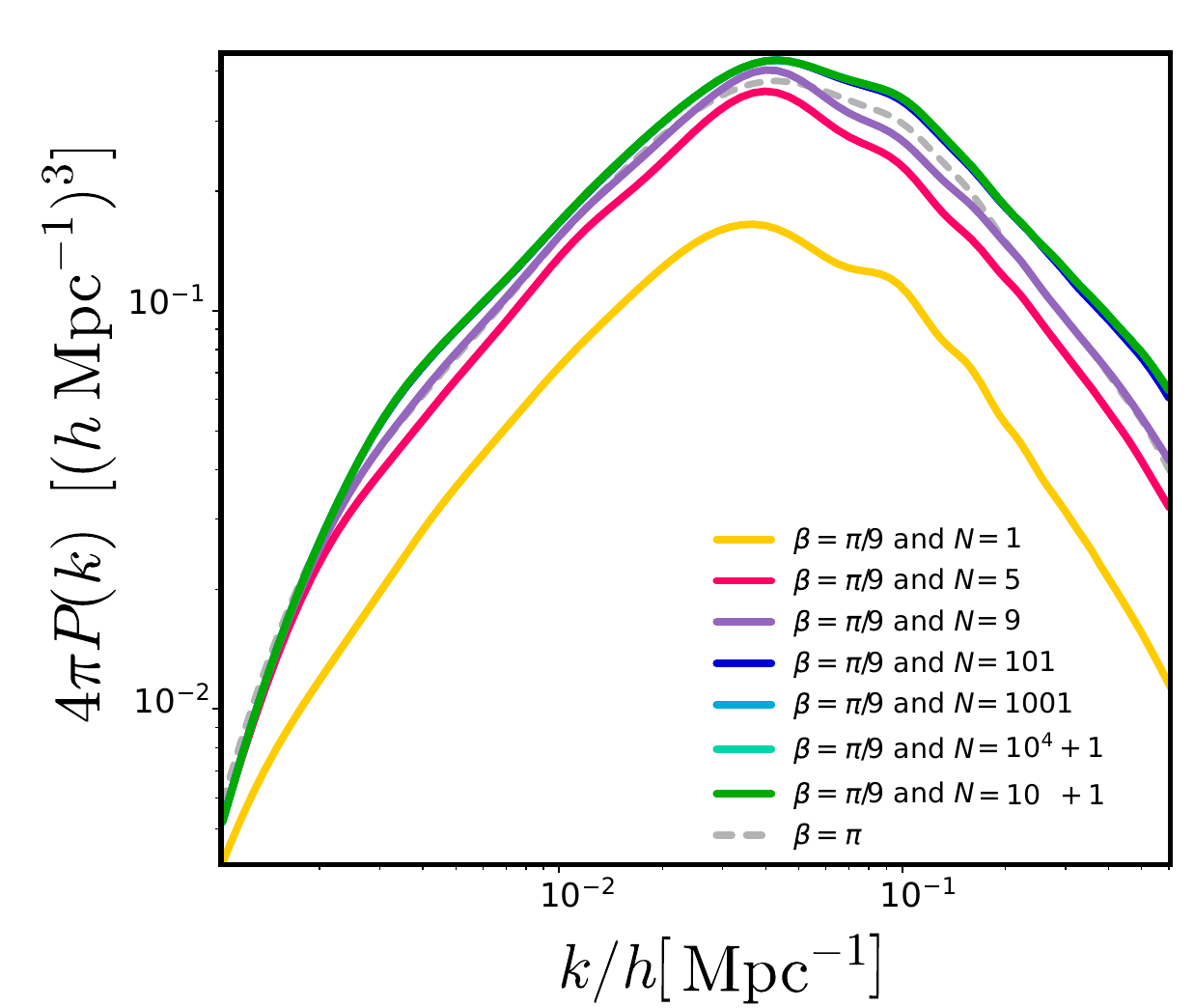}
    \caption{CDM Linear power spectrum generated by wiggly cosmic strings for different values of $N$ and $\beta=\pi/9$. The dashed line corresponds to the power spectrum generated by straight cosmic strings. Here we set $G\mu_{\rm eff}=10^{-7}$ and the results are obtained by averaging over 500 realizations.}
    \label{fig:PSkink}
\end{figure}

\begin{widetext}

\begin{figure}[h!] 
            \centering
            \includegraphics[width=7in]{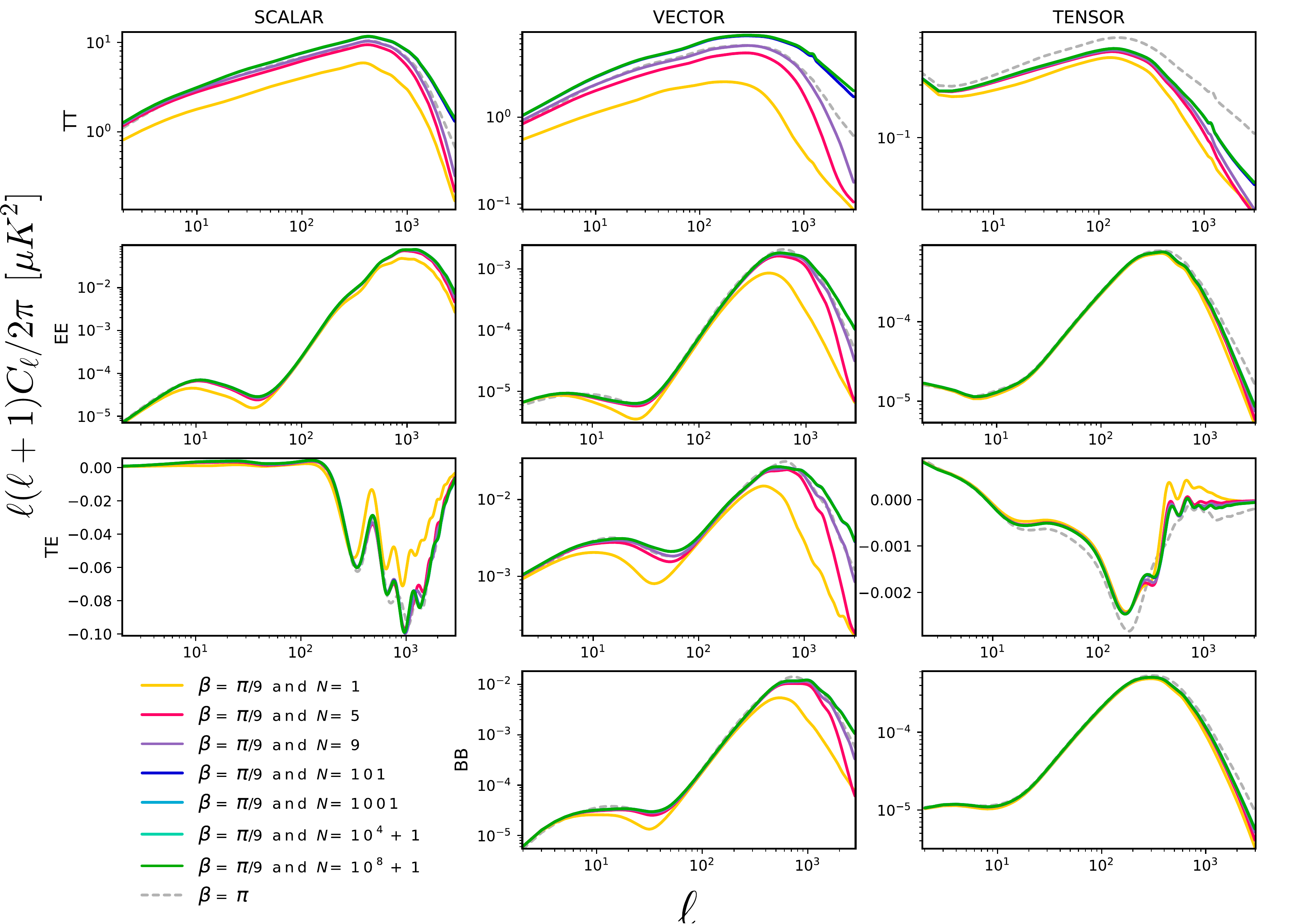}
            \caption{CMB anisotropies generated by wiggly cosmic strings for different values of $N$ and $\beta=\pi/9$. From top to bottom, we plot the TT, EE, TE and BB power spectra, as a function of the multipole moment $\ell$. The left, middle and right columns represent the scalar, vector and tensor components, respectively. The dashed lines correspond to the angular power spectra generated by straight cosmic strings. Here we set $G\mu_{\rm eff}=10^{-7}$ and the results are obtained by averaging over 500 realizations.}
            \label{fig:CMBkink}
\end{figure}   

\end{widetext}

Let us now study the impact of the number of kinks per segment on the CMB anisotropies. Here, we will fix $\beta$ to $\pi/9$ for illustration purposes, but we have extensively studied the spectra for different values of kink sharpness. We will briefly comment on what happens for other values of $\beta$, in the cases in which different behaviour was observed.

In Fig.~\ref{fig:PSkink}, we plot the linear CDM power spectrum for $\beta=\pi/9$ and different values of $N$, alongside the spectrum generated by straight string segments. This figure shows that increasing the number of kinks per segment leads to an increase of the height of the peak of the spectrum, which is accompanied by its broadening and a shift towards smaller scales. This figure also shows, however, that if the number of kinks is increased beyond a given value, the shape is roughly maintained. This behaviour may, again, be explained by the fact that, by changing $N$, we are changing the length of the straight subsegments --- in fact, $d\propto 1/(N+1)$ --- thus shifting again power from large to small-scales. In this case, however, we are also increasing the number of subsegments, thus adding power to the corresponding scales. For $\beta=\pi/9$, we have that $d \sim 6 L/(N+1)$ and thus, for small enough $N$, the length of the subsegment is comparable to $L$ and this power is added to a scale close to the peak. However, as we further increase $N$, $d\ll L$ and, as a consequence, the spectrum is affected on increasingly smaller scales (or larger $k$). For large enough $N$, these scales are, in fact, beyond the range of the plot. For smaller sharpness, we see precisely the same type of behaviour, but this saturation of the spectrum happens for smaller values of $N$. We may thus say that increasing the number of kinks generally leads to an enhancement of the spectrum up to scales comparable to the interkink distance.

A similar behaviour may be seen in the CMB anisotropies generated by wiggly strings as $N$ is increased, as displayed in the angular power spectra plotted in Fig.~\ref{fig:CMBkink}. As a matter of fact, there is, generally, an enhancement of the amplitude of the angular power spectra, specially for larger multipole moment $\ell$. This enhancement, in the case of the scalar and vector components of the temperature anisotropies, spans the whole range of the power spectrum and may be quite significant in the case of the vector modes --- which further supports the idea that the pointy geometry of the kink favours the generation of vortical motions of matter on all scales. The behaviour of the tensor component of the TT anisotropies is however more complex. In this case, for $\beta=\pi/9$, the increase is mostly seen at smaller scales than in the spectra of the other components (nevertheless for large $\ell$ it may still be clearly seen). We found, however, that if the kinks are less sharp, the amplitude of the spectrum on large scales may actually decrease with increasing $N$. The spectrum, nevertheless, hints at enhancements at very small-scales, which seems to indicate that the tensor components are affected by the kinks at smaller scales (larger $\ell$) than the scalar and vector counterparts. As we decrease segment length, by increasing $\beta$, these changes move towards larger and larger $\ell$ and, as matter of fact, significant power is shifted from large to small-scales. We found that this change in behaviour happens roughly around $\beta\sim\pi/4$, for which $d\sim 2.6/(N+1)$, and $d<L$ for small values of $N$.

The behaviour of the E-mode and B-mode polarization angular power spectra is qualitatively similar to that of the TT anisotropies. There is an increase of the amplitude of the spectrum for large enough $\ell$ in all cases, but this is generally slight in the case of the scalar EE spectrum. This enhancement is more prominent and covers a wider range of $\ell$ in the case of the vector modes; in fact, the peaks of the spectra are both broader and shifted slightly towards higher $\ell$. As seems to be the case for the TT anisotropies, the tensor component of the E-mode and B-mode polarization are only affected at larger $\ell.$

\subsection{Comparison with other models}\label{subsec:comp}

The CMB anisotropies generated by wiggly cosmic strings were originally studied in~\cite{Pogosian:1999np} by resorting to the effective model to describe wiggly strings introduced in~\cite{Carter:1990nb,Vilenkin:1990mz}. The basic premise of this effective model is that a long string with small-scale structure appears to be smooth to an observer who cannot resolve this structure, but --- unlike ``bare'' strings --- it also appears to have a mass per unit length $\mu_{\rm eff}$ and a tension $T_{\rm eff}$ that are different and distinct from $\mu_0$. As a matter of fact, these strings appear to be heavier by a factor of $\alpha$ and to have a lower tension:

\begin{equation}
    \mu_{\rm eff}=\alpha \mu_0\quad\mbox{and}\quad T_{\rm eff}=\frac{\mu_0}{\alpha}\,,
    \label{eq:eos}
\end{equation}
where $\alpha\ge 1$ is a parameter that controls how much heavier these strings are. Although in this effective model $\alpha$ is not related to the geometry of the segments, by comparing with Eq.~(\ref{eq:mueff}), one may see that this parameter is akin to our wiggliness parameter $\mathcal{S}$, which is related to the sharpness of kinks. In this effective approach string segments are still treated as straight, however, their equation of state (\ref{eq:eos}) is  different from that of standard strings. Wiggly strings then behave, in this effective model, as a superposition of a straight string --- but with an angular deficit that is enhanced by a factor of $\alpha$ --- with a linelike distribution of matter~\cite{Vachaspati:1991tt}. There is therefore an additional gravitational force on particles, and the strings accrete matter.

The impact of wiggliness on the CMB anisotropies, in this effective model, may be more easily understood by looking at the stress-energy tensor for such a string~\cite{Pogosian:1999np}:

\begin{equation}
    \Theta_{ij}=\left[v^2\hat{\dot{X}}_i{\hat{\dot{X}}}_j - \frac{\left(1 - v^{2} \right)}{\alpha^2}\hat{{X}}^{\prime}_{i} {\hat{{X}}^{\prime}}_{j}\right]\Theta_{00}\,,
    \label{eq:Tij-eff}
\end{equation}
where $\Theta_{00}$ is the Fourier transform of the temporal component of the stress-energy tensor of a straight cosmic string (which may be obtained from Eq.~(\ref{T00}) by setting $\beta=\pi$ \footnote{Apart from the factor of $\alpha^2$, Eq.~(\ref{eq:Tij-eff}) can also be obtained from Eq.~(\ref{Tij}) by setting $\beta=\pi$.}). The impact of wiggliness then is to suppress the $\Theta_{ij}$ components with respect to $\Theta_{00}$ and this suppression leads to a relative decrease of the vector and tensor mode contributions relative to the scalar modes~\cite{Pogosian:2007gi}. This is similar to what we found with our ``geometric'' approach in Sec.~\ref{subsec:sharp} for $\beta<\pi/2$. As a matter of fact, for a `zig-zag' segment, the stress components in Eq.~(\ref{Tij}) may be rewritten as

\begin{figure}[h!]
    \centering
    \includegraphics[width=3.4in]{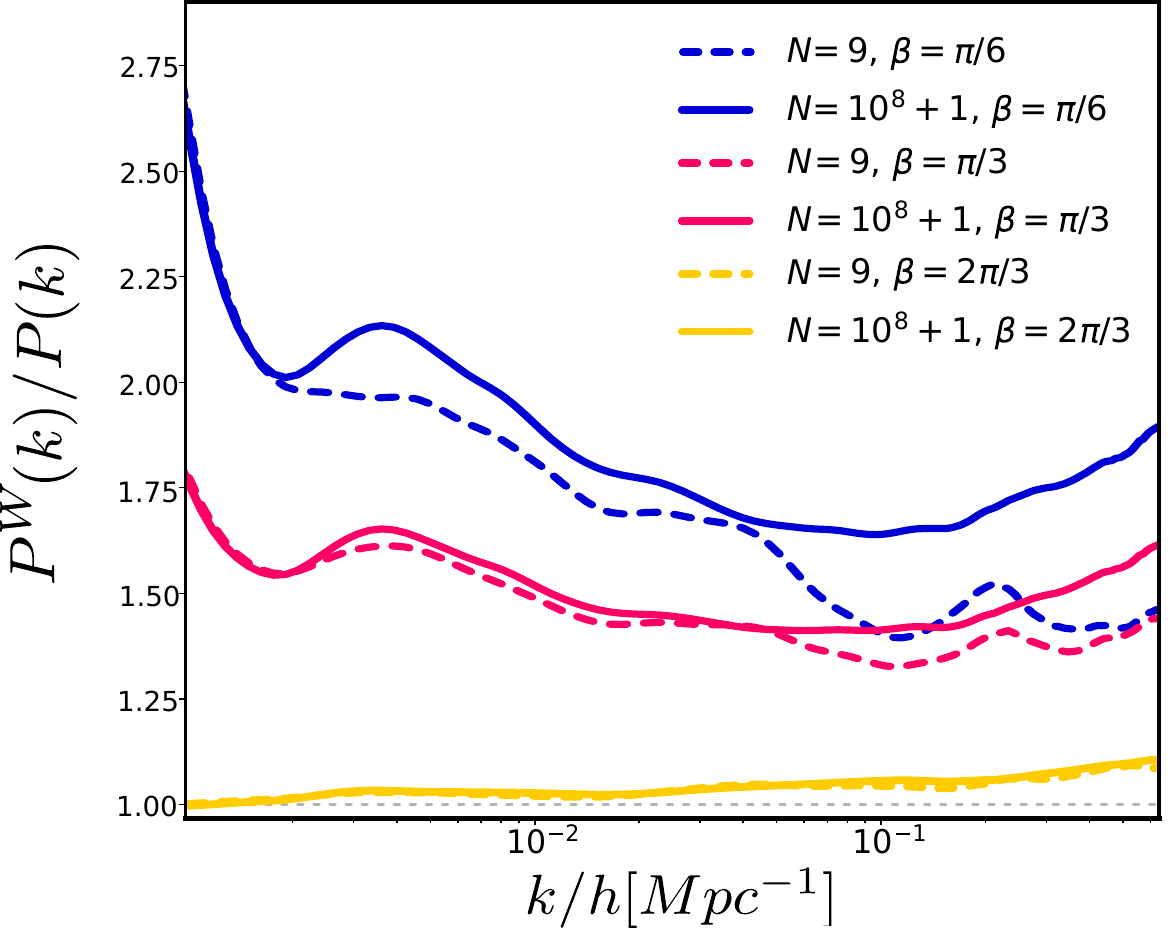}
    \caption{Ratio between CDM Linear power spectrum generated by wiggly cosmic string segments for different values of sharpness and that of the effective model with the same renormalized mass per unit length $\mu_{\rm eff}$. Solid lines correspond to $N=10^8+1$, while dashed lines represent $N=9$. The results are obtained by averaging over 500 realizations.}
    \label{fig:Comp1}
\end{figure}

\begin{eqnarray}
     \frac{\Theta_{ij}}{\Theta_{00}} & = & v^2\hat{\dot{X}}_i{\hat{\dot{X}}}_j - \left(1 - v^{2} \right)\left[\hat{{X}}^{\prime}_{1i} {\hat{{X}}^{\prime}}_{1j}\right. \\
     & - & \left.\sin{\beta}\left(\hat{e}_{x\,i}\hat{e}_{y\,j}+\hat{e}_{x\,j}\hat{e}_{y\,i}\right)\frac{\Theta_{00\,(2)}}{\Theta_{00}}\right]\nonumber\,,
\end{eqnarray}
which shows that, in this case, there is also a reduction of the same term of $\Theta_{ij}$ relative to $\Theta_{00}$. Writing $\Theta_{ij}$ in this form also shows that, for these `zig-zag segments', this reduction is maximum for $\beta=\pi/2$\footnote{Apart, off course, for the dependence of $\Theta_{00\,(2)}/{\Theta_{00}}$ on $\beta$.}, which explains why the amplitude of the vector and tensor modes start to increase if $\beta$ decreases below $\pi/2$. This behaviour for very sharp kinks --- or equivalent large $\mu_{\rm eff}$ --- is not predicted in the effective model and, thus, in general, we should expect this model to predict smaller vector and tensor anisotropies in this limit.

\begin{widetext}

\begin{figure}[h!] 
            \centering
            \includegraphics[width=7in]{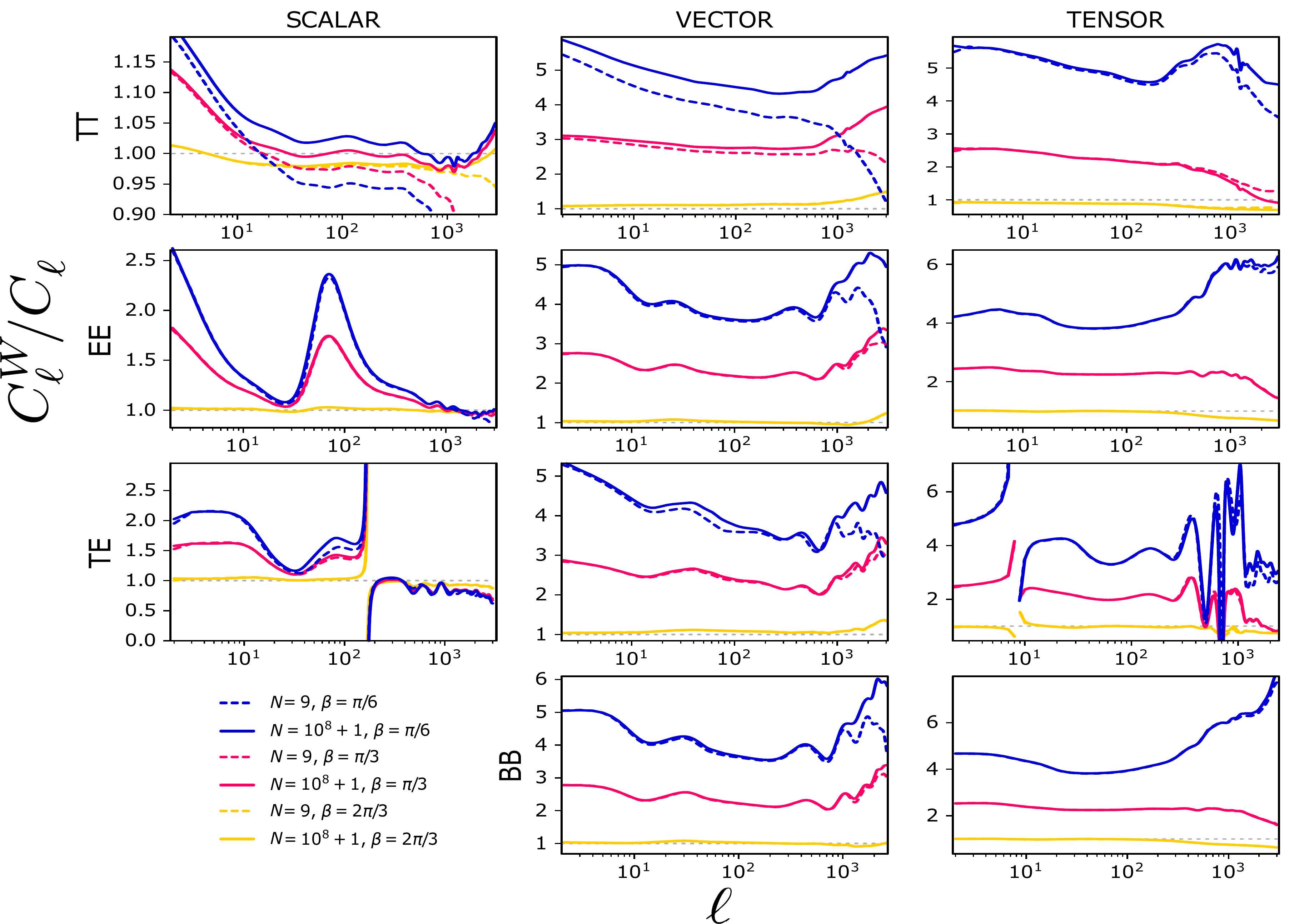}
            \caption{Ratio between CMB angular power spectrum generated by wiggly cosmic string segments for different values of sharpness and those predicted by the effective model with the same renormalized mass per unit length $\mu_{\rm eff}$. Solid lines correspond to $N=10^8+1$, while dashed lines represent $N=9$. From top to bottom, we plot the TT, EE, TE and BB power spectra, as a function of the multipole moment $\ell$. The left, middle and right columns represent the scalar, vector and tensor components, respectively. The dashed lines correspond to the angular power spectra generated by straight cosmic strings. The results are obtained by averaging over 500 realizations.}
            \label{fig:Comp2}
\end{figure}   

\end{widetext}

In Fig.~\ref{fig:Comp1}, we plot the ratio between the CDM linear power spectrum generated by cosmic strings with kinks and that of effectively wiggly strings with the same renormalized mass per unit length (i.e, with $\alpha=\mathcal{S}$). This figure shows that, indeed, as we increase $\mu_{\rm eff}$, the predictions of these two approaches start to diverge, with the effective model predicting systematically lower amplitudes. The same behaviour may be seen in Fig.~\ref{fig:Comp2}, where we plot the ratio between the CMB anisotropies produced by both models. As before, the predictions of the effective model seem to differ the most from those of a network of strings with kinks when $\beta$ is the furthest from $\pi$. In the case of the scalar contribution to the temperature anisotropies, quite generally there is an underestimation of the amplitude of the spectrum at large scales, which in the large $N$ limit is accompanied also by an underestimation at small-scales of about a few percent. For intermediate scales, around $10<\ell<10^3$, this underestimation becomes less significant and, for small enough sharpness and/or small $N$, there is an overestimation instead. Regarding the scalar E-mode polarization, our geometrical approach predicts a higher and broader reionization peak (at $\ell\sim 10$) and a broader peak at $\ell=10^3$. This difference again increases with increasing $\beta$, although it is not strongly dependent on $N$.

Qualitatively, the differences in the vector and tensor contributions to the temperature anisotropies and to the E-mode and B-mode polarization spectra are very similar. Vector modes generally are underestimated over all scales in the effective approach, which may indicate that including the kinks explicitly may be necessary to accurately predict these contributions. This underestimation is generally more significant for large multipole moments and for larger $N$. For the tensor components, however, provided that sharpness is not too large, the predictions of the effective model actually exceed those of the model introduced here for large enough $\ell$. However, as one increases sharpness, there is an inversion of this behaviour and, as was the case for the vector modes, the effective model actually underestimates the amplitude of the spectrum by increasingly large amounts. This increase in the difference between the predictions of the different models as we increase $\beta$ is probably related to the change in behaviour for vector and tensor components when $\beta$ decreases below $\pi/2$.

Finally, it is worth noting that, although the effective model should, in general, be expected to provide a better description in the limit in which $d$ is very small --- i. e. if the distance between kinks is so small that an observer cannot resolve it --- here we found that the results are more discrepant in the large $N$ limit in almost every case.

\section{Discussion and conclusions}
\label{Conc}

Here, we have developed an alternative formalism to describe the CMB anisotropies generated by wiggly cosmic strings. The novel feature of our model is that, instead of modeling wiggly strings effectively by considering straight segments with a modified equation of state, we have considered explicitly piecewise straight segments with kinks. As a matter of fact, we have derived a closed-form expression for the stress-energy tensor of these zig-zag segments that has two free parameters: kink sharpness and the distance between kinks. This has then allowed us to extend the USM to describe wiggly segments and to use CMBACT to study the anisotropies generated by cosmic strings with small-scale structure for a large variety of parameters.

Our results show that, at large scales, the temperature anisotropies are mostly determined by the energy density of the network. As matter of fact, we have verified numerically that for a multipole moment of $\ell=10$ --- which is commonly used to derive constraints on the cosmic string mass per unit length $\mu_0$ --- there are no significant deviations from the scaling relation in Eq.~(\ref{eq:mueff-sca}). As a result, constraints on $G\mu_0$ derived using CMB data can straightforwardly be generalized into constraints on $G\mu_{\rm eff}$. Constraints on the cosmic string mass per unit length, then, are more stringent as sharpness increases:
\begin{equation}
G\mu_{0}  \mathcal{S} < 1.49\times 10^{-7} \, ,
\label{eq:cons}
\end{equation}
where we have used the constraints obtained in~\cite{Lazanu:2014eya}. Note however that this does not describe the full picture: when we consider zig-zag strings, we are, in fact, introducing a new scale to the problem --- the interkink distance --- and this in general causes an enhancement of the anisotropies beyond that predicted in Eq.~(\ref{eq:mueff-sca}) on the multipole range corresponding to this scale. As matter of fact, we generally found an increase of the temperature anisotropies on sufficiently small-scales, especially in the large $N$ limit. This is mostly due to a significant enhancement of the vector modes that is caused by the discontinuities on the string tangent at the tip of the kink --- which are not accounted for in the effective model. This enhancement of vector modes at large $\ell$ is quite a general feature of the angular power spectra generated by these wiggly string segments. The anisotropies generated by cosmic strings are in general expected to dominate the CMB angular power spectra at very small-scales, since --- unlike inflationary perturbations --- they evade Silk-damping by actively generating anisotropies after last scattering~\cite{Pogosian:2008am}. This additional contribution from kinks at small-scales --- together with the potential contribution of cosmic string loops studied in~\cite{Rybak:2021scp} --- may further exacerbate the dominance of strings at small-scales.

Note that, in our approach, the impact of small-scale structure is included purely at a geometrical level and we did not consider its impact on the cosmic string dynamics. On the one hand, kinks are expected to propagate along the strings and small-scale structure is expected to oscillate relativistically. Whether the inclusion of the effects of this microscopic motion will make the predictions of the model developed here and that of the effective model more alike or not is an open question that warrants further investigation. Nevertheless, it seems reasonable to expect that some differences will remain due to the fact that we are introducing an additional scale to the problem and that kinks themselves seem to have a significant impact on the anisotropies. Moreover, the cosmological evolution of cosmic string networks may also be affected by the existence of small-scale structure on cosmic strings, which in a sense adds an additional inertia that slows the networks at large scales. As a matter of fact, the impact of small-scale structure on network dynamics was considered in \cite{Martins:2014kda, PhysRevD.94.096005, Rybak:2017yfu} in the context of the effective model and the authors found that this may result in a decrease in the amplitude of the CMB anisotropies. However, these studies do not take the geometrical impact of kinks into consideration in the computation of the CMB power spectra and they necessarily underestimate the impact of wiggliness as a result. 

Moreover, here we have assumed that the sharpness of the kinks and the number of kinks per string remains fixed throughout the cosmological evolution of the network. Kinks are, however, created dynamically as a result of string interactions and intercommutations and, when these result in the production of loops, some kinks may be removed too. Moreover, kinks are also subject to being stretched by expansion and to decay due to gravitational backreaction. These phenomena change the average distance between kinks and their sharpness as the network evolves and this may have an impact on the CMB anisotropies as well. The model we propose here, however, may be used straightforwardly with an evolving kink sharpness and number of kinks per segment to also account for the dynamical effects of small-scale structure, for example, by using more complex models to describe the large-scale dynamics of wiggly networks that exist in the literature such as~\cite{Vilenkin:1990mz, Austin:1993, Carter:1994zs, Carter:1997pb, Martins:2014kda}.

\acknowledgments

The authors would like to thank Pedro P. Avelino for illuminating discussions and Huangyu Xiao for drawing our attention to a mistake in Eq.~(\ref{eq:cons}). L. S. is supported by FCT - Funda\c{c}\~{a}o para a Ci\^{e}ncia e a Tecnologia through contract No. DL 57/2016/CP1364/CT0001. This work was also supported by FCT through the research grants UIDB/04434/2020 and UIDP/04434/2020 and through the R \& D project 2022.03495.PTDC - \textit{Uncovering the nature of cosmic strings}. This work was also financed by FEDER---Fundo Europeu de Desenvolvimento Regional funds through the COMPETE 2020---Operational Programme for Competitiveness and Internationalisation (POCI), and by Portuguese funds through FCT in the framework of the project POCI-01-0145-FEDER-031938 and PTDC/FIS-PAR/31938/2017.

\bibliography{Biblio}

\end{document}